\documentclass[a4paper,11pt]{article}
\pdfoutput=1 % if your are submitting a pdflatex (i.e. if you have
\usepackage{jheppub} % for details on the use of the package, please
\usepackage[T1]{fontenc} % if needed

\usepackage{color}
\usepackage{caption}
\usepackage{subcaption}
\usepackage{amsmath}
\usepackage{amssymb}
\usepackage{amsthm}
\usepackage{physics}
\usepackage{dsfont, mathtools}
\usepackage{bm}
\usepackage{url}
\usepackage{setspace}

\usepackage{float}

\title{A quantum circuit interpretation of\\
evaporating black hole geometry}

\author{Ying Zhao}
\affiliation{Institute for Advanced Study, Princeton, NJ 08540, USA}

\emailAdd{zhaoying@ias.edu}

\usepackage{color}

\abstract{We give a quantum circuit interpretation of evaporating black hole geometry. We make an analogy between the appearance of island for evaporating black hole and the transition from two-sided to one-sided black hole in the familiar example of perturbed thermofield double. If Alice perturbs thermofield double and waits for scrambling time, she will have a one-sided black hole with interior of her own. We argue that by similar mechanism the radiation gets access to the interior (island forms) after Page time. The growth of the island happens as a result of the constant transitions from two-sided to one-sided black holes. 

}

\begin{document} 
\maketitle
\flushbottom

\section{Introduction}

In the recent work of \cite{Penington:2019npb} and \cite{Almheiri:2019psf}, the authors studied the geometry of an evaporating black hole. They found that there is a jump of the location of RT surface around Page time. In this paper we give an interpretation of this geometry from the point of view of quantum circuit.  

Consider an evaporating black hole formed from a pure state as discussed in \cite{Penington:2019npb,Almheiri:2019psf}. Before Page time, the RT surface is empty set and the interior is inside the entanglement wedge of the black hole. After Page time, the interior is mostly inside the entanglement wedge of the radiation due to the appearance of an island \cite{Almheiri:2019hni}.

This may seem puzzling from the point of view of radiation since just after the transition we have a large jump in the size of the region contained in its entanglement wedge. Why is it that adding a few more qubits to the radiation gives rise to such large change? Why does the bulk region belonging to the radiation (island in \cite{Almheiri:2019hni}) keep growing after Page time while the radiation itself doesn't do any computations? In this paper we will answer these questions from the point of view of quantum circuit. 

For a holographic state with a bulk dual, the bulk geometry reflects the minimal circuit preparing the state \cite{Swingle:2012wq} \cite{Hartman:2013qma} \cite{Susskind:2014moa}. From studying the uncomplexity of subsystems, we identify different parts of the quantum circuit as belonging to different subsystems \cite{Zhao:2017isy}.  

In section \ref{quantum_circuit}, we briefly review some ideas about quantum circuit we use in this paper. In particular, we will identify different parts of the quantum circuit as belonging to different subsystems. We revisit the familiar example of perturbed thermofield double and look at the transition from two-sided to one-sided black hole\footnote{Here, when we say Alice has a one-sided black hole at some time, we mean that if she jumps into the black hole at that time, her experience can be completely determined by the operations on her side. We will discuss this in more detail later.} after scrambling. In section \ref{Evaporating_BH} we apply these ideas to evaporating black holes. We will see that the appearance of the island  belonging to the radiation \cite{Almheiri:2019hni} can also be understood as a transition from two-sided to one-sided black hole. As evaporation continues, such transitions constantly happen and as a result the island keeps growing.  

The difference is that, for perturbed thermofield double, the scrambling is a result of future time evolution. For evaporating black hole, the scrambling is done by the existing circuit stored in the wormhole on a spatial slice.

\section{Review of quantum circuit}
\label{quantum_circuit}

\subsection{Bulk tensor network and quantum circuit}

In \cite{Swingle:2012wq}, Swingle argued that AdS bulk geometry reflects a MERA-like tensor network preparing the CFT ground state. In \cite{Hartman:2013qma}, Hartman and Maldacena made a connection between tensor network and time evolution. They pointed out that for thermofield double, the Hamiltonian evolution at thermal scale make up the tensor network corresponding to the interior geometry.

\begin{figure}[H] 
 \begin{center}                      
      \includegraphics[width=2in]{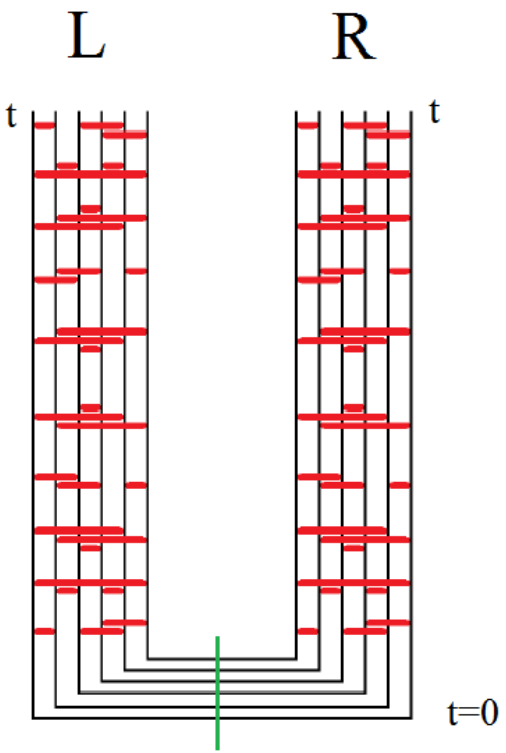}
      \caption{A quantum circuit representing the time evolution of thermofield double.}
  \label{picture1}
  \end{center}
\end{figure}

\begin{figure}[H] 
 \begin{center}                      
      \includegraphics[width=3.6in]{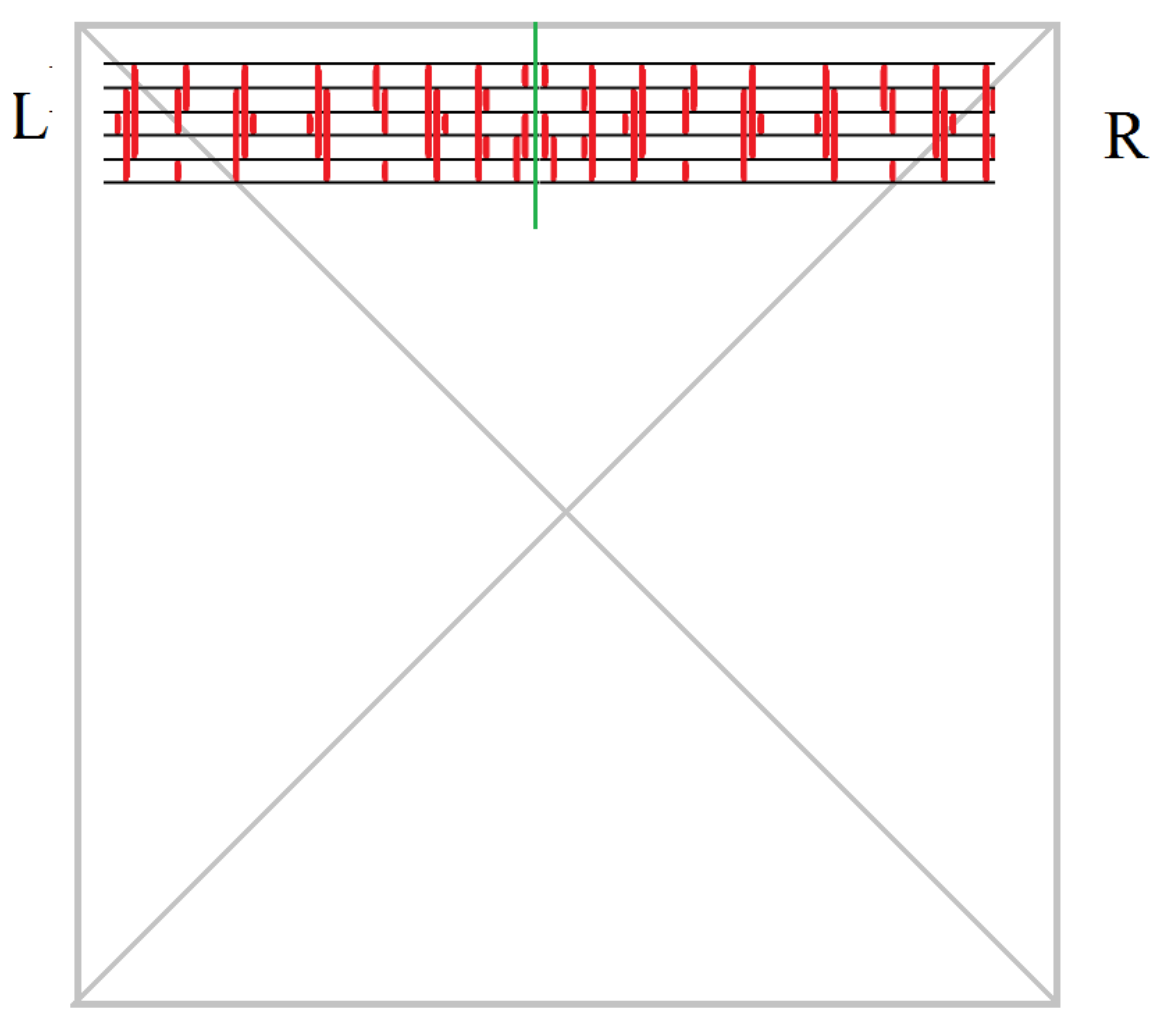}
      \caption{Quantum circuiti in Figure \ref{picture1} becomes the tensor network representing ERB at time $t$.}
  \label{picture2}
  \end{center}
\end{figure}

In \cite{Susskind:2014moa}, Susskind described a quantum circuit model of black hole time evolution. Figure \ref{picture1} illustrates the evolution of thermofield double.\footnote{This figure is from \cite{Susskind:2014jwa}. I thank Leonard Susskind for allowing me to use this figure.} The initial state is a product of Bell pairs and the red lines represent gates on two qubits.
Susskind pointed out that once we lift the quantum circuit in Figure \ref{picture1} and lay it on a spatial slice, we get a tensor network description of wormhole at time $t$. Figure \ref{picture2} is also from \cite{Susskind:2014moa}. The gates in the quantum circuit of Figure \ref{picture1} become tensors in the tensor network representing the wormhole (ERB) at time $t$. In what follows, we will say that the gates are stored in the wormhole. 

\subsection{Quantum circuit belonging to different subsystems}

When the system is composed of two parts A and B, we can identify part of the quantum circuit as belonging to subsystem A and part of the circuit as belonging to subsystem B. 

As argued in \cite{Susskind:2015toa}, the growth of the black hole interior is fueled by the uncomplexity of a state \cite{Brown:2017jil}. When the system is composed of two parts: A and B, we can look at the uncomplexity of each subsystem \cite{Zhao:2017isy}.

Assume A and B together are in an entangled state $\ket{\psi}$. We focus on subsystem A. As long as the state $\ket{\psi}$ is not maximally complex, generic operations on A will decrease the uncomplexity of $\ket{\psi}$ \cite{Brown:2017jil}. Among these operations, some of them do not change the density matrix $\rho_A$. If we write $\rho_A$ in Schmidt basis, it's easy to see that the unitary operations that do not change $\rho_A$ are the relative rotations of the Schmidt basis. They can be undone from subsystem B. These operations only make sense when we have both subsystems A and B. They do not belong to either subsystem alone as they do not change the density matrix of A or B. 

On the other hand, there are unitary operations on A that cannot be undone from subsystem B. They will contribute to A's uncomplexity. Consequently, when there is a quantum circuit connecting A and B, we identify those gates that can be undone from A but not from B as belonging to subsystem A, those gates that can be undone from B but not from A as belonging to subsystem B.

This has a holographic interpretation. Consider combined AB system in a pure state $\ket{\psi}$ with a dual bulk geometry. The bulk geometry reflects the minimal tensor network preparing this state. In \cite{Zhao:2017isy} we made the following observations related to subregion duality \cite{Dong:2016eik}. 

\begin{figure}[H] 
 \begin{center}                      
      \includegraphics[width=2in]{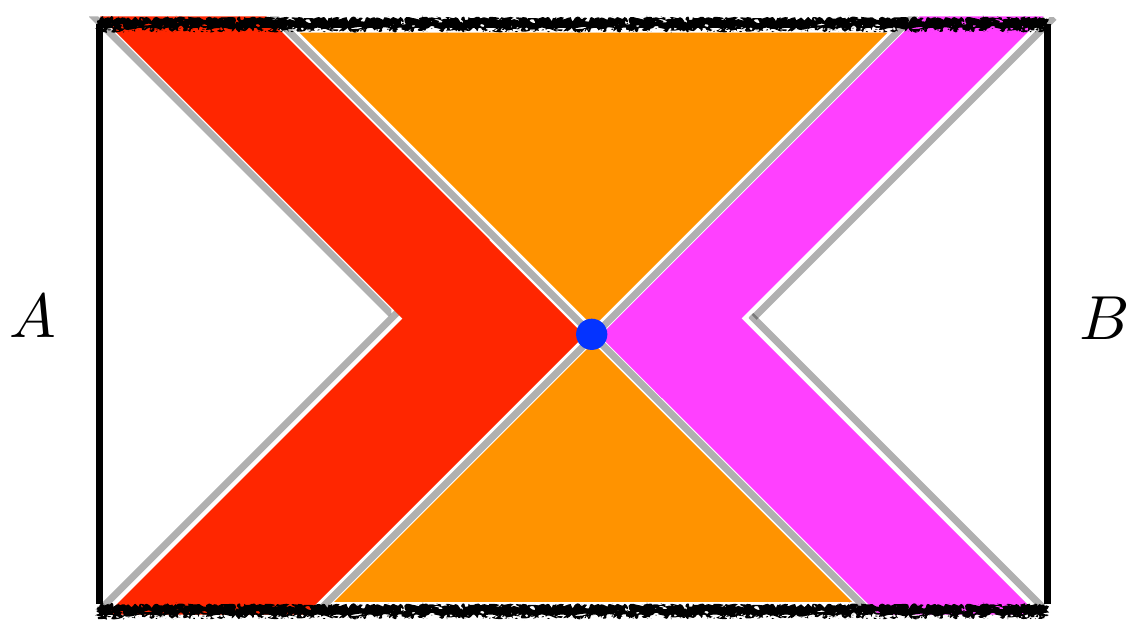}
      \caption{}
  \label{uncomplexityspacetime}
  \end{center}
\end{figure}

In Figure \ref{uncomplexityspacetime}, the blue dot represents the RT surface whose area gives the entanglement entropy between subsystems A and B. The gates in the tensor network that can be undone from A but not from B are stored inside A's entanglement wedge (red region in Figure \ref{uncomplexityspacetime}). The gates that can be undone from B but not from A are stored in B's entanglement wedge (pink region in Figure \ref{uncomplexityspacetime}). The gates that can be undone from both sides do not belong to either subsystem. They are stored in the entanglement region which is out of both entanglement wedges (orange region in Figure \ref{uncomplexityspacetime}). 

In this paper we will study evaporating black hole from these observations. 

\subsection{Transition from two-sided to one-sided black hole}

\subsubsection{Two-sided black hole}

We start from thermofield double shared between Alice and Bob. Alice has subsystem A and Bob has subsystem B. In terms of quantum circuit Alice and Bob share $S$ Bell pairs. 
At this stage they have zero uncomplexity as their computations can be undone by each other. Figure \ref{thermofield_double} shows the corresponding quantum circuit and wormhole geometry at $t_L = t_R =0$. The wormhole is shortest at this time. 

\begin{figure}[H] 
 \begin{center}                      
      \includegraphics[width=5in]{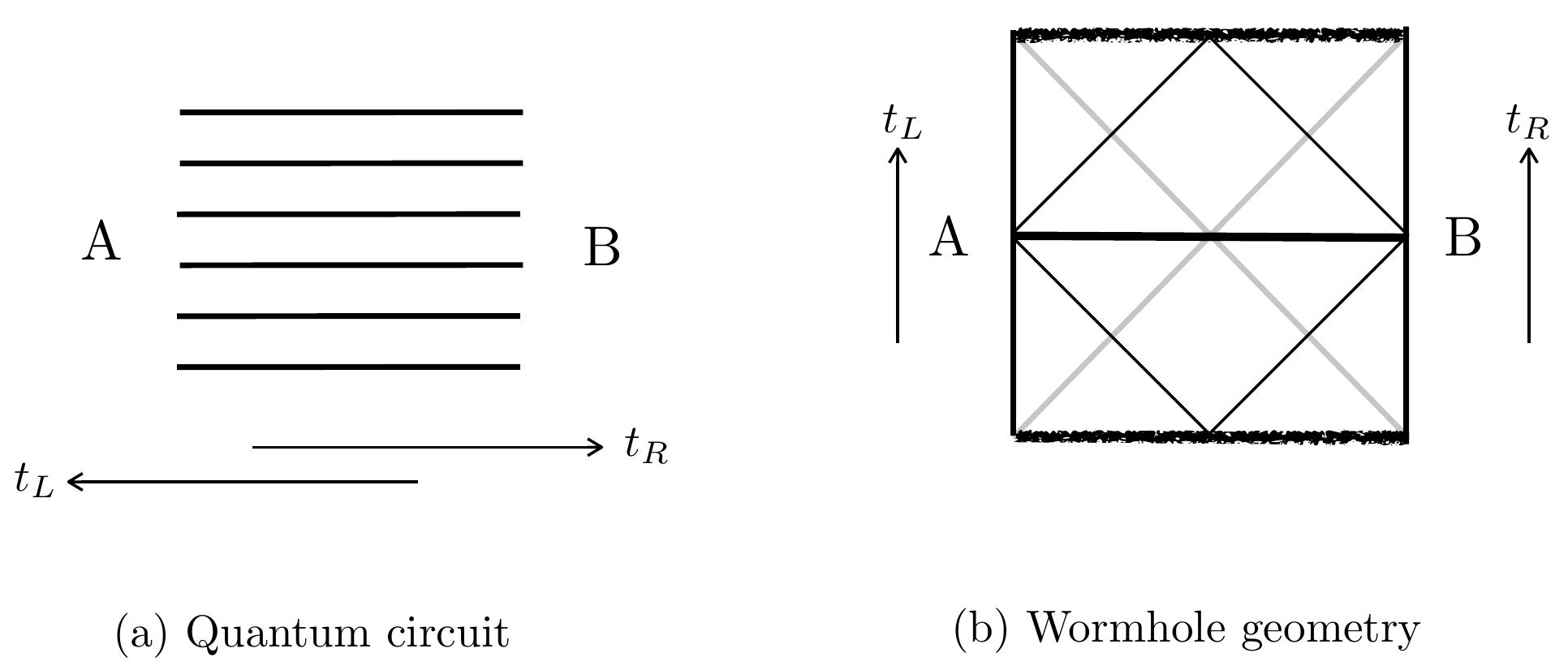}
      \caption{We represent thermofield double by $S$ Bell pairs.}
    \label{thermofield_double}
  \label{thermofield_double}
  \end{center}
\end{figure}

\begin{figure}[H] 
 \begin{center}                      
      \includegraphics[width=5in]{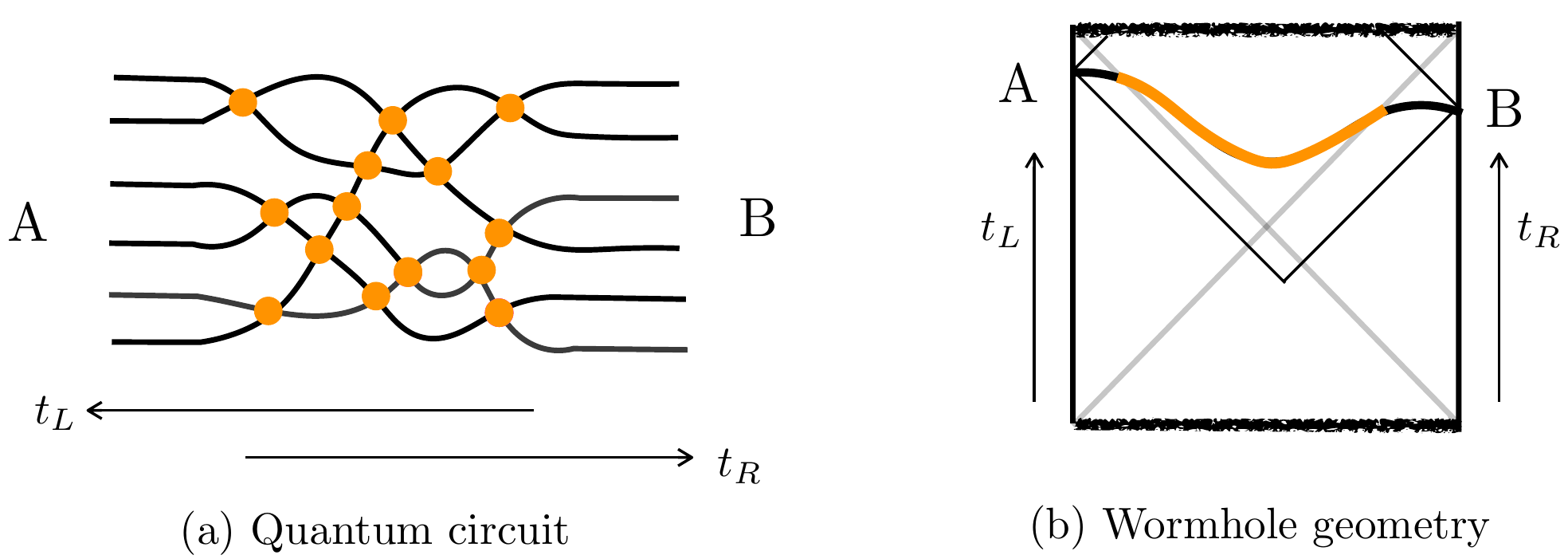}
      \caption{Time-evolved thermofield double}
    \label{thermofield_double_time_evolved}
  \end{center}
\end{figure}

As we increases $t_L$ or $t_R$, the minimal quantum circuit preparing the state gets longer ((a) in Figure \ref{thermofield_double_time_evolved}). The circuit is stored on the wormhole and the wormhole also gets longer ((b) in Figure \ref{thermofield_double_time_evolved}). Notice that the gates stored in the quantum circuit can be undone from both A and B. Correspondingly, the wormhole can be shortened from both A and B. The interior region (orange line in Figure \ref{thermofield_double_time_evolved}) does not belong to either A and B. It belongs to the union and A and B by the miracle of ER = EPR \cite{Maldacena:2013xja}.

\subsubsection{Compare two-sided and one-sided black holes}

When Alice has a two-sided black hole, the interior is outside her entanglement wedge. Alice cannot predict an infalling observer's experience. From the point of view of the quantum circuit, as the circuit gets longer, the gates in the circuit does not change Alice's density matrix. The expansion of the interior is fueled by the relative rotation of the Schmidt basis, not Alice's own uncomplexity.

When Alice has a one-sided black hole, she can predict an infalling observer's experience. The expansion of the interior is fueled by Alice's uncomplexity. Because we use the term of ``one-sided black hole" in such a flexible way, the property of a black hole being two-sided or one-sided can be a time-dependent quantity. In particular, we do not mean counting the number of boundaries. We will come back to this later.

\subsubsection{Perturbed thermofield double}

The bulk geometry of thermofield double with a simple perturbation was studied in \cite{Shenker:2013pqa}. The authors found that there will be significant backreaction after scrambling time. We look at it from the point of view of quantum circuit \cite{Zhao:2017isy}. In the next section we will draw analogy with the appearance and growth of island in case of an evaporating black hole.

We start form thermofield double shared by Alice and Bob. Now Alice throws in $c$ qubits from the left side at time $t_w$ where $c$ is much less than the black hole entropy. In \cite{Susskind:2014jwa}\cite{Brown:2016wib}, an epidemic model was used to study perturbed quantum circuit and the growth of a precursor. We briefly review it here. Represent the black hole dynamics by Hayden-Preskill type circuit \cite{Hayden:2007cs}: At each time step (every thermal time) the qubits are randomly grouped into $\frac{S}{2}$ pairs, and on each pair a randomly chosen $2$-qubit gate is applied. We can characterize the effect of some small perturbation in such a system as follows. Imagine the unpertubed system contains $S$ healthy qubits, and the perturbation is $c$ extra qubits carrying some disease. The sick qubits enter the system at $t = 0$. Any qubits who interact directly or indirectly with sick qubits will get sick. We define the size of the epidemic $s(t)$ to be the number of sick qubits at time $\tau$. It grows exponentially in time and saturates at scrambling time. With $\log S$ large, $\frac{s(\tau)}{S}$ is almost like a step function. 

During each thermal time, $S$ gates are applied.\footnote{We count each 2-qubit gate as $2$ gates.} Among them, $s(\tau)$ gates are affected by the sick qubits while $S-s(\tau)$ gates remain healthy. As Bob does not have access to the extra sick qubits, he can undo those healthy gates but he cannot unto those sick gates.

\begin{figure}[H] 
 \begin{center}                      
      \includegraphics[width=4in]{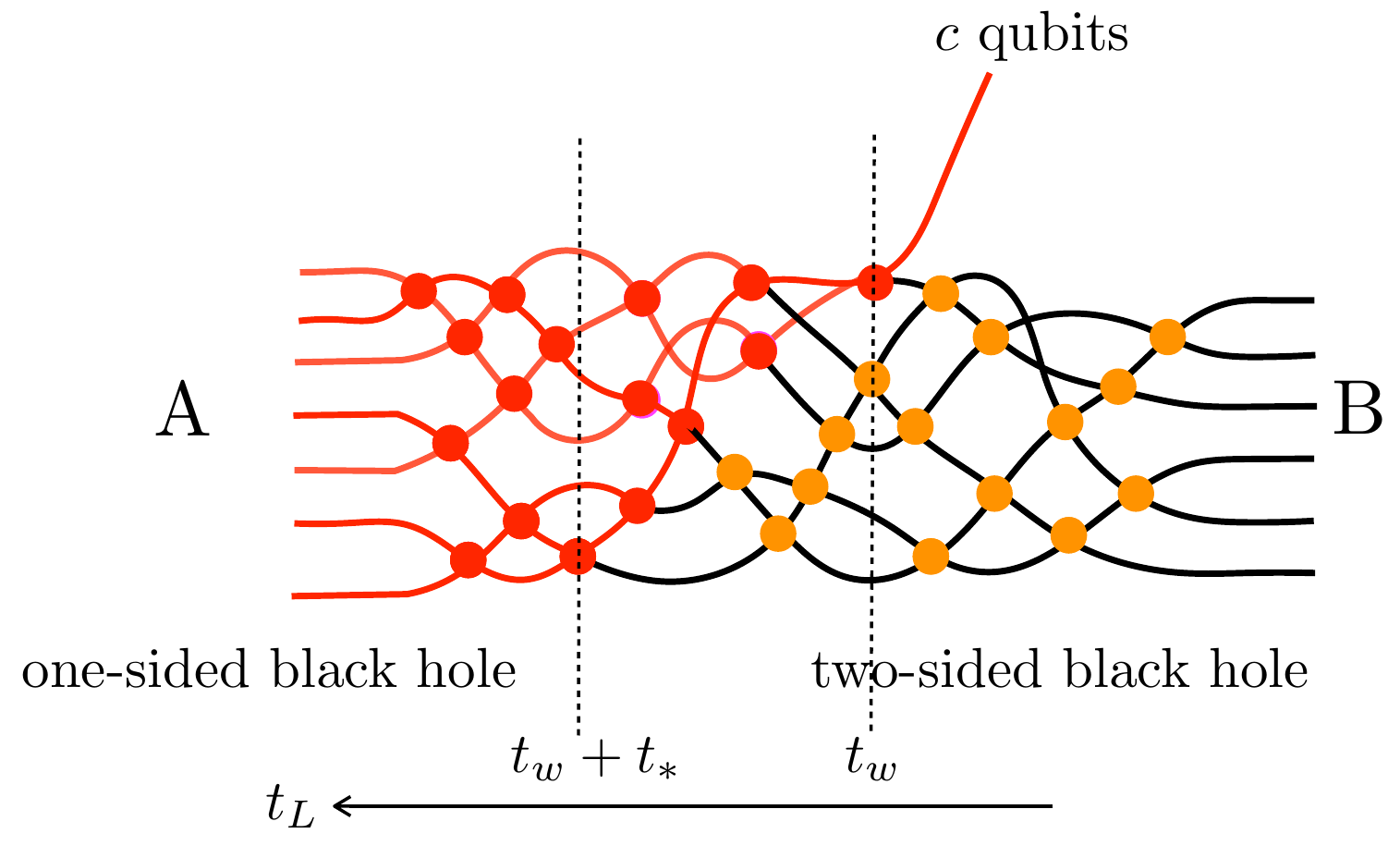}
      \caption{Quantum circuit representing perturbed thermofield double}
    \label{thermofield_double_perturbed_circuit}
  \end{center}
\end{figure}

\begin{figure}[H] 
 \begin{center}                      
      \includegraphics[width=6in]{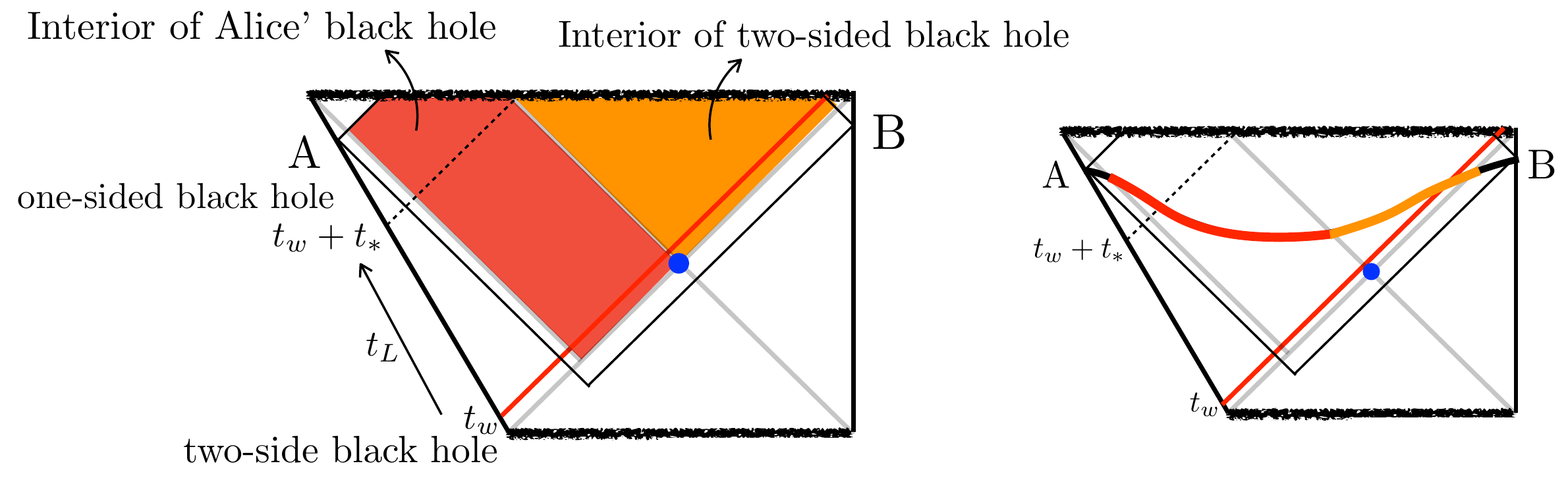}
      \caption{Perturbed thermofield double. The blue dot is the RT surface. The orange region is the interior of a two-sided black hole and depends on both A and B. The red region is the interior of Alice's black hole.}
  \label{thermofield_double_perturbed_Penrose}
  \end{center}
\end{figure}

Figure \ref{thermofield_double_perturbed_circuit} shows the corresponding quantum circuit as we increase Alice's time $t_L$. Such a quantum circuit is stored on the wormhole (Figure \ref{thermofield_double_perturbed_Penrose}). When the perturbation just comes in at time $t_w$, it has little effect on Alice's system. For $t_w<t_L<t_w+t_*$, most of the qubits in subsystem A are not affected by the perturbation and the quantum gates on them (orange dots in Figure \ref{thermofield_double_perturbed_circuit}) can still be undone from subsystem B. These gates shared by Alice and Bob are stored in the orange region in Figure \ref{thermofield_double_perturbed_Penrose}. It is outside both entanglement wedges and it is the interior of a two-sided black hole. 
In this time window Alice essentially still has a two-sided black hole, i.e., the interior accessible to her can be affected by Bob. At $t_w+t_*$ all her qubits are affected by the perturbation. After that when $t_L>t_w+t_*$, the quantum gates applied on her side (red dots in Figure \ref{thermofield_double_perturbed_circuit}) can no longer be undone from Bob side. These quantum gates belonging to Alice are stored in the red region in Figure \ref{thermofield_double_perturbed_Penrose}. That region is inside Alice's entanglement wedge. For a more detailed comparison between the quantum circuit analysis and the growth of the wormhole geometry, see \cite{Zhao:2017isy}. 

When $t_L>t_w+t_*$ Alice's black hole becomes effectively one sided because the portion of the interior accessible to her is determined only by operations on her subsystem. In what follows we will refer to this situation by saying Alice's black hole has become one-sided. The red region is the interior of her one-sided black hole.

In this example, the perturbation is scrambled by Alice's future time evolution and there is a transition from two-sided black hole for one-sided black hole for Alice.\footnote{In appendix \ref{charged} we examine perturbed thermofield double of charged black holes. I thank Edward Witten for raising up the potential problem with charged black holes.}

Before we move on to evaporating black hole, we make some comments on this qubit model. Here we model thermofield double by maximally entangled state. Without the extra qubit, the action of any unitary on A can be undone by B. These unitaries correspond to the Hamiltonian time evolution for finite temperature thermofield double. For finite temperature thermofield double, the gates that can be undone by B are phases on Schmidt basis, which is energy basis. 

For density matrix $\rho_A = \frac{1}{Z}e^{-\beta H}$, the coarse-grained entropy equals the fine-grained entropy. Any operations that cannot be undone from subsystem B will change this density matrix. It will increase the coarse-grained entropy such that the coarse-grained entropy will be larger than the entanglement entropy. That's why we model such operations by introducing extra qubits.

\section{Evaporating black hole}
\label{Evaporating_BH}

For the convenience of discussion, in this section we assume Bob has the remaining black hole while Alice has the radiation. There is one obvious difference between the case of evaporating black hole and the previous example of perturbed thermofield double. For evaporating black hole, all the computations are done on Bob's black hole side and there are no computations done on Alice side. Despite of this, Alice still gets her share of interior, essentially by the same mechanism as in the previous example.

\subsection{Before Page time: Bob's one-sided black hole}

Before Page time, the radiation consists of less than half of the entire system and it is maximally entangled with the black hole. Bob can reduce the state to $S_{BH}$ Bell pairs while Alice cannot reduce the complexity of the state. 

From our earlier assumptions, the gates stored in the interior belong to Bob. Bob has his own one-sided black hole and the wormhole storing the gates is his interior.

\subsection{Around Page time: two-sided black hole}
 
\begin{figure}[H] 
 \begin{center}                      
      \includegraphics[width=4in]{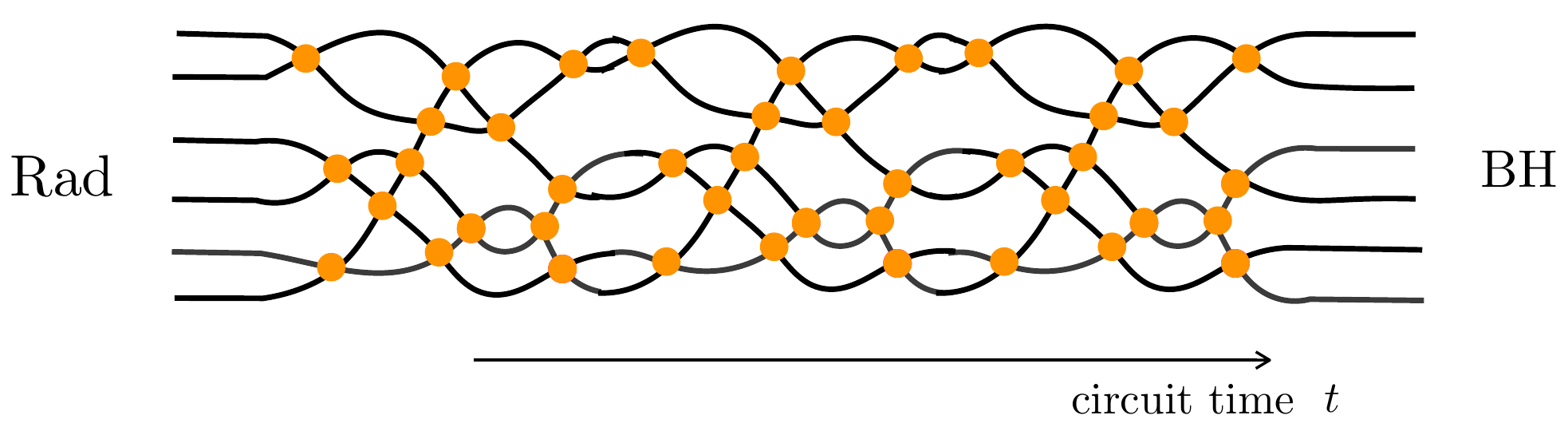}
      \caption{}
  \label{Page_time}
  \end{center}
\end{figure}

At around Page time, the black hole and the radiation are maximally entangled. There is a wormhole with complexity $\sim S^2$ \footnote{One needs to be careful about the definition of complexity here. See \cite{Toappear}.} connecting the black hole and the radiation. See the circuit picture in Figure \ref{Page_time}. 

Unlike the case before Page time, the quantum circuit does not belong to Bob's subsystem alone. The gates can be undone by Alice and they belong to the union of the radiation and the black hole. 

\begin{figure}[H] 
 \begin{center}                      
      \includegraphics[width=2.6in]{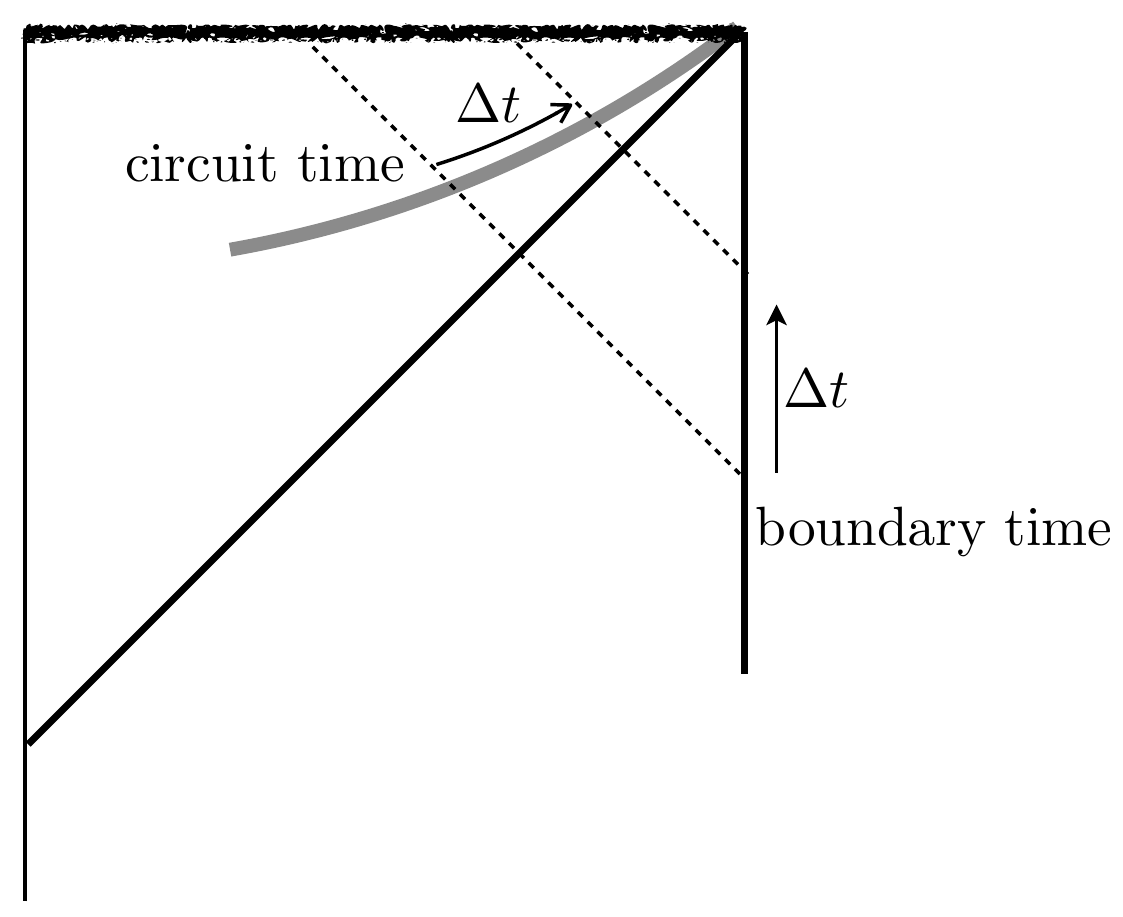}
      \caption{}
  \label{symmetry_across_horizon}
  \end{center}
\end{figure}

Note that the circuit time going right corresponds to the radial location of the wormhole. Since all the computations are done by Bob, there is an identification between Bob's time and the interior radial location, so we can also identify the circuit time with Bob's time (Figure \ref{symmetry_across_horizon}).\cite{Zhao:2017iul} We will use this fact later.

\subsection{After Page time}

\begin{figure}[H] 
 \begin{center}                      
      \includegraphics[width=6in]{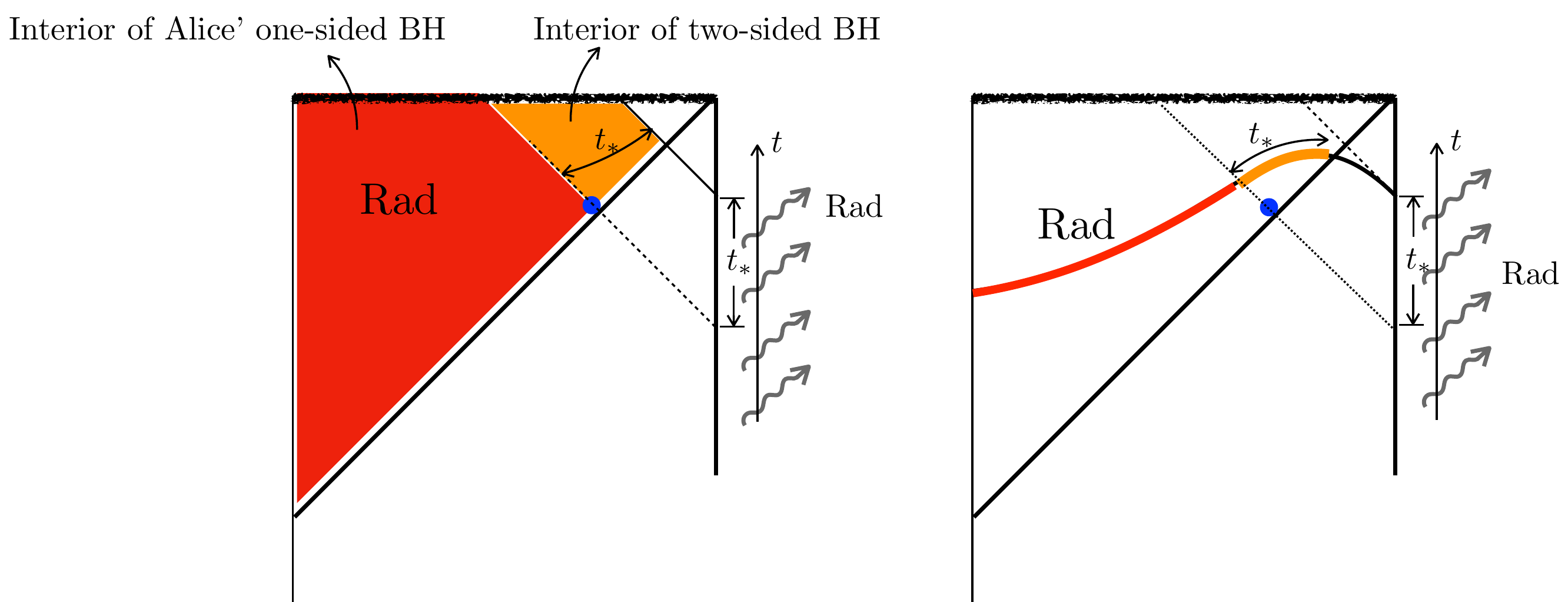}
      \caption{The blue dot represents the RT surface.}
  \label{After_Page_Penrose}
  \end{center}
\end{figure}

After Page time, the black hole is maximally entangled with the radiation so Bob's computations can be undone from the radiation by Alice. From this point of view, Bob's remaining black hole is a two-sided black hole. This is consistent with the result in \cite{Penington:2019npb, Almheiri:2019psf} that the RT surface (blue dot in Figure \ref{After_Page_Penrose}) lies almost at the horizon, i.e., Bob's remaining black hole does not have its own interior. If he jumps into the black hole after Page time, he will enter the entanglement region (orange region in Figure \ref{After_Page_Penrose}) and Alice can affect his experience in the interior. This is the resolution of firewall paradox given in \cite{Maldacena:2013xja}.

As Bob's time increases, the quantum circuit grows and new gates will be stored in the entanglement region. The orange region in Figure \ref{After_Page_Penrose} grows from the right end. On the other hand, the black hole also gives some of its qubits to the radiation. Let's look at its effect from the point of view of quantum circuit. 

\begin{figure}[H] 
 \begin{center}                      
      \includegraphics[width=4in]{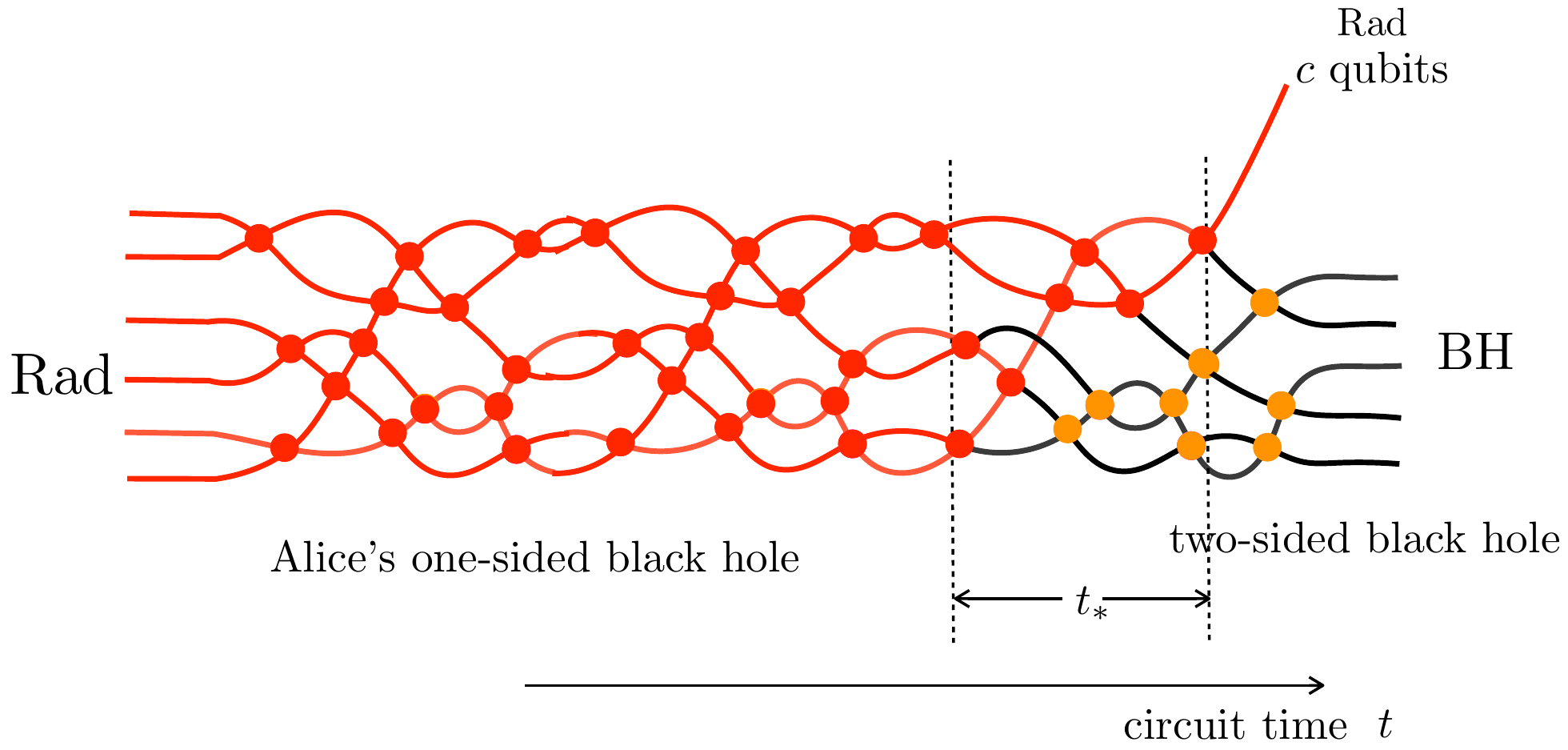}
      \caption{}
  \label{After_Page_circuit}
  \end{center}
\end{figure}

We start from a quantum circuit connecting Alice's radiation and Bob's black hole as in Figure \ref{Page_time}. It stores Bob's computations until Page time. Now Bob takes out $c$ qubits and gives them to Alice. We follow these $c$ qubits in the quantum circuit and go backward in the circuit time, i.e., we read Figure \ref{After_Page_circuit} from the right to the left. At the right end of the circuit there are $S_{BH}-c$ qubits belonging to Bob's black hole. Then $c$ qubits belonging to Alice come in. After circuit time $t_*$, the extra $c$ qubits completely scramble the circuits and any gates on the left of that (red dots in Figure \ref{After_Page_circuit}) can no longer be undone by Bob. They belong to Alice. Alice gets her share of the circuit despite that she didn't do any of these computations. 

The quantum circuit in Figure \ref{After_Page_circuit} lies on the wormhole in Figure \ref{After_Page_Penrose}. The red gates in Figure \ref{After_Page_circuit} form the island in \cite{Almheiri:2019hni}. The interior geometry belonging to the radiation corresponds to the part of the quantum circuit that cannot be undone from the black hole side.

We showed that if we trace the quantum circuit backward by scrambling time, we encounter the gates belonging to Alice. This scrambling time is horizontal circuit time. We mentioned before that there is identification between the circuit time and Bob's boundary time. As a result the Eddington-Finkelstein time of the RT surface also stays scrambling time behind (Figure \ref{After_Page_Penrose}) as shown in \cite{Penington:2019npb, Almheiri:2019psf}.

\subsubsection{Comments on one-sided black hole}

Notice that unlike in the example of perturbed thermofield double, the radiation system does not have a classical horizon. Then why do we say that the island is like the interior of a one-sided black hole? We note the similarity between the quantum circuits in Figure \ref{After_Page_circuit} and Figure \ref{thermofield_double_perturbed_circuit}. In both figures, we have a minimal quantum circuit preparing the pure state of combined AB system. In a circuit like this, the gates that can be undone from both A and B form the interior of two-sided black hole, i.e., the part of spacetime region outside both A's entanglement wedge and B's entanglement wedge. The gates that can be undone from side A but not from side B belongs to A. That's why we say the red gates in Figure \ref{After_Page_circuit} form the interior of Alice's one-sided black hole. 

There is another question. One can say that the red gates belong to Alice's entanglement wedge, but why do we say Alice has a one-sided black hole? This has to do with the nature of the quantum circuit. Consider an ordinary one-sided black hole formed from collapse, and let's ask what's the difference between the interior and exterior spacetime region.  

The bulk geometry reflects the minimal quantum circuit preparing the state. The part of the circuit that corresponds to the exterior geometry is a RG circuit. To learn about this part of circuit, all one needs to know are some thermodynamic coarse-grained properties of the state, for example, mass, charge, temperature of the black hole, e.t.c. On the gravity side, one can use HKLL reconstruction to reconstruct the exterior operators\cite{Hamilton:2005ju}. 

On the other hand, the part of the circuit corresponding to the interior geometry is an unitary quantum circuit. This part of the circuit encodes fine-grained knowledge about the state, and it is genetically very complex to figure out\cite{Bouland:2019pvu}. On the gravity side, one can use Petz map to reconstruct the interior operators but it's based on the prior knowledge of the code subspace and the reconstruction is state-dependent\cite{Penington:2019kki}.

Now let's look at the circuit connecting the radiation to the black hole. It was argued that the state of the radiation is pseudorandom, i.e., with simple operations one cannot tell the difference between the state of the radiation and maximally mixed state\cite{Kim:2020cds}. In particular, with simple operations one can not figure out the red gates in Figure \ref{After_Page_circuit}. So that part of the circuit encodes fine-grained information about the radiation system and is like the interior of a black hole. \footnote{There is another way to look at this. For an ordinary one-sided black hole formed from collapse, the RG circuit corresponding to the exterior takes the total number of qubits in the field theory (with some cutoff) down to the coarse-grained entropy at the horizon. For the radiation system in the qubit model, the coarse-grained entropy equals to the total number of qubits in the radiation, so there is no need for RG circuit giving rise to the exterior and the quantum circuit belonging it corresponds to its interior.}

\subsection{Growth of the island: Transition from two-sided to one-sided black holes constantly happening}

\begin{figure}[H] 
 \begin{center}                      
      \includegraphics[width=5in]{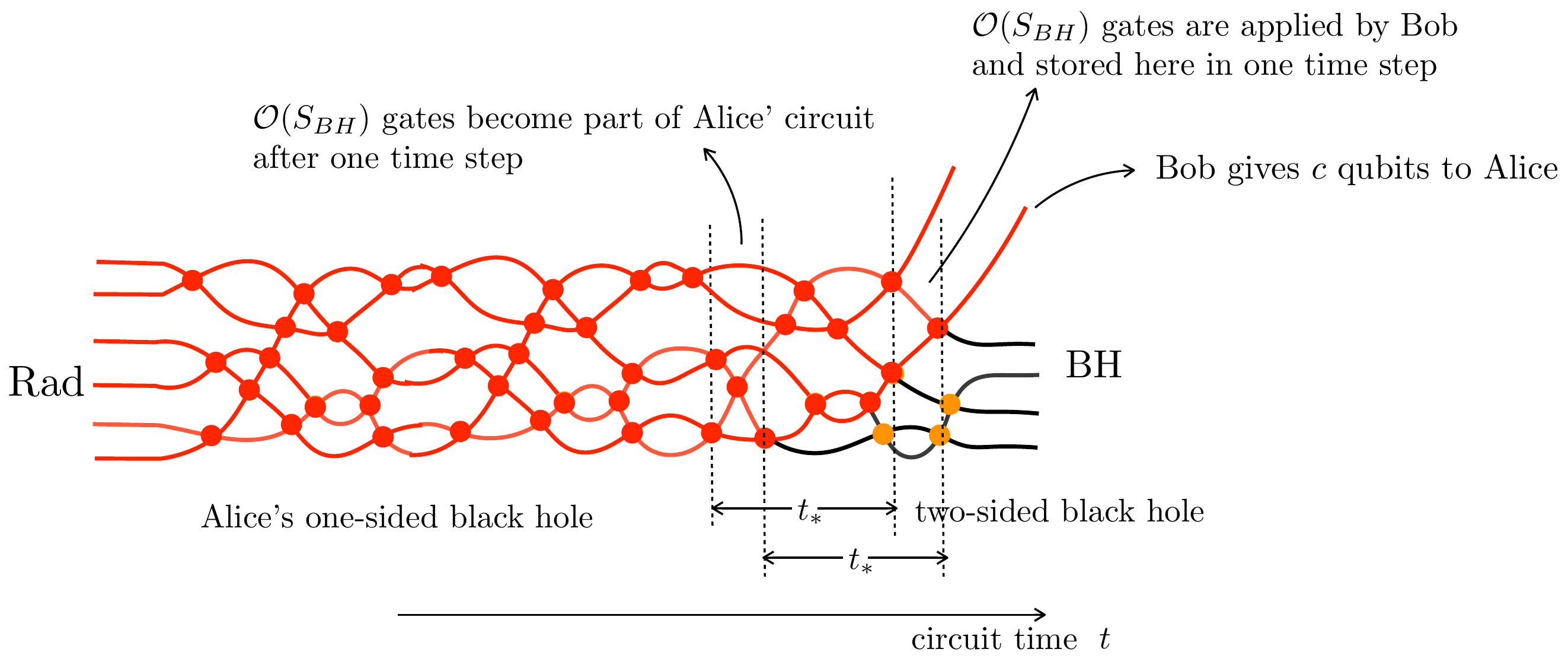}
      \caption{}
  \label{After_Page_circuit_2}
  \end{center}
\end{figure}

After another time step (thermal time), from Bob's time evolution $\mathcal{O}(S_{BH})$ gates are applied and stored on the circuit . Another $c$ qubits go to the radiation. The part of the circuits affected by these $c$ qubits become Alice's circuit. See figure \ref{After_Page_circuit_2}.

\begin{figure}[H] 
 \begin{center}                      
      \includegraphics[width=5in]{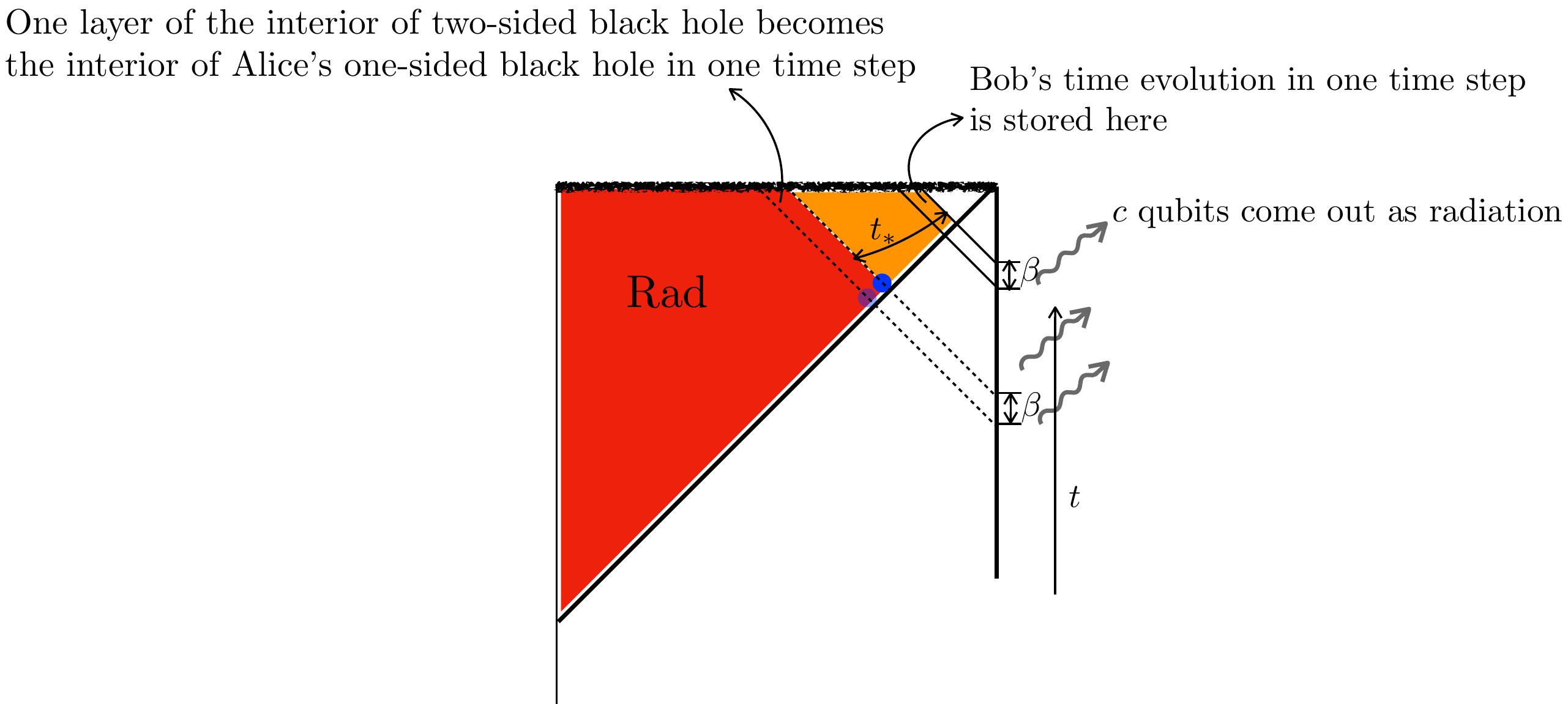}
      \caption{The blue dot represents the RT surface. It moves along the horizon and always stays scrambling time behind.}
  \label{After_Page_Penrose_2}
  \end{center}
\end{figure}

Correspondingly, the island belonging to the radiation grows and the RT surface moves outward along the horizon. It always stays scrambling time behind \cite{Penington:2019npb,Almheiri:2019psf}. Figure \ref{After_Page_Penrose_2} shows what happens to the geometry after one time step. As the radiation process continues the transitions from two-sided to one-sided black hole constantly happen. The growth of the island is a result of such transitions.

Notice that the quantum circuits in Figure \ref{thermofield_double_perturbed_circuit} and Figure \ref{After_Page_circuit} look the same. Figure \ref{thermofield_double_perturbed_circuit} represents Alice's time evolution while figure \ref{After_Page_circuit} represents the quantum circuit stored in the wormhole on a spatial slice. 

For the case of perturbed thermofield double, the extra qubits enter Alice's side and the scrambling computations are done on her side. After the extra qubits completely scramble her subsystem,  Alice's computations completely belong to herself and Alice has a one-sided black hole. 

In the case of evaporating black hole after Page time, Bob's black hole does all the computations and stores the circuit on the wormhole connected to the radiation. When he gives some of his qubits to Alice, this existing circuit stored on the wormhole will scramble these qubits and part of the circuit becomes Alice's interior despite that she didn't do any of the computations. In this sense, the appearance and growth of the island belonging to the radiation is a transition from two-sided to one-sided black hole just like in the familiar example of perturbed thermofield double.

 \section*{Acknowledgments}

I am grateful to Leonard Susskind for helpful comments on the draft and for allowing me to use figures from his paper. I thank Ahmed Almheiri, Adam Levine, Raghu Mahajan, Juan Maldacena, Geoff Penington, Leonard Susskind, and Edward Witten for helpful discussions. I thank the referee for helpful comments and questions. I am supported by the Simons foundation through the It from Qubit Collaboration.

\appendix

\section{Charged black hole geometry}
\label{charged}

\begin{figure}[H] 
 \begin{center}                      
      \includegraphics[width=4in]{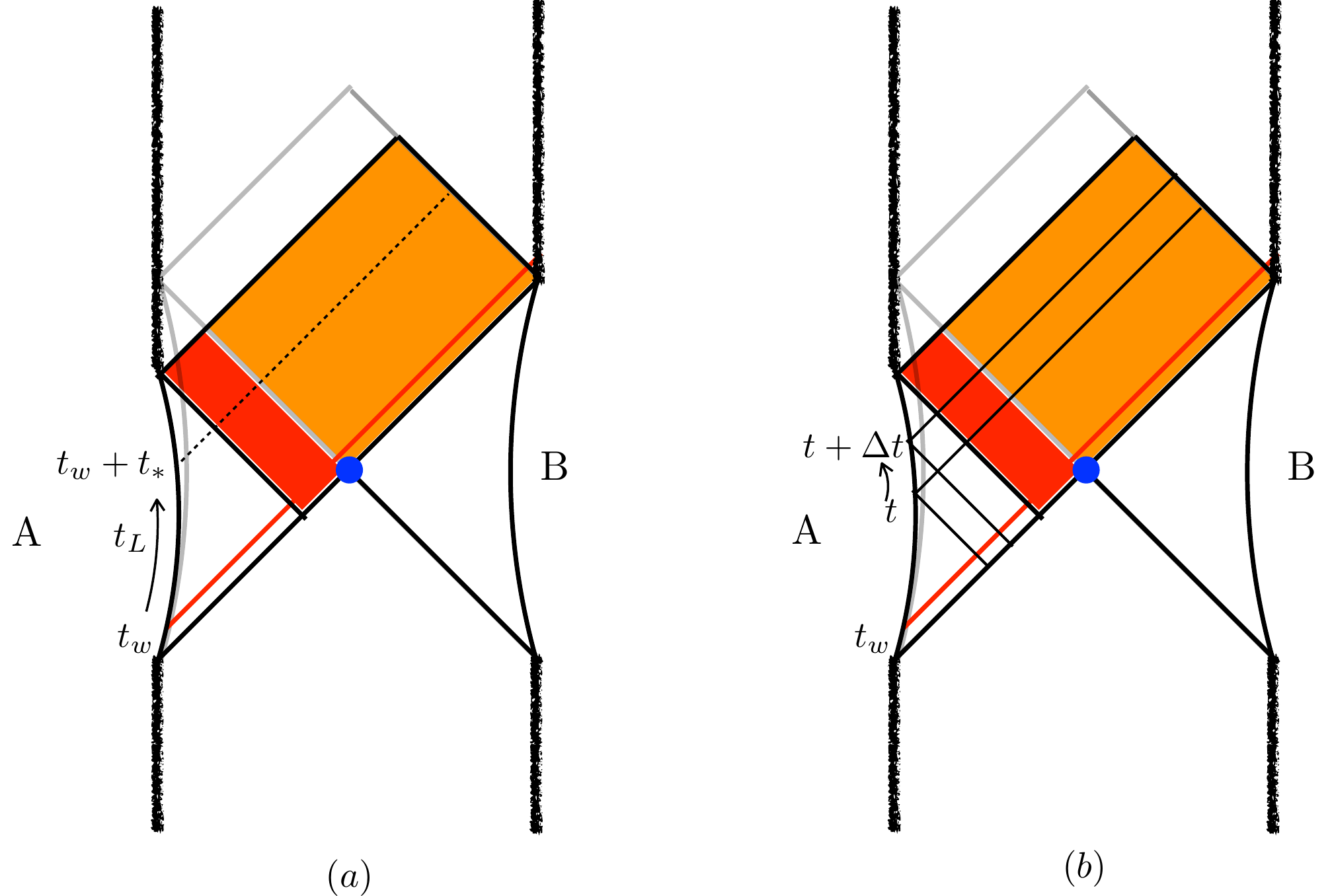}
      \caption{}
  \label{Charged_BH_perturbed}
  \end{center}
\end{figure}

In this paper, we used the property that if one perturbs two-sided black hole and waits for scrambling time, it will become a one-sided black hole. However, if we start from thermofield double of charged black holes, from the Penrose diagram it looks like no matter how late Alice waits after throwing in the perturbation, Bob can always affect her from the right side (Figure \ref{Charged_BH_perturbed} (a)). Naively this seems to imply that in this case Alice never gets her one-sided black hole. We study the geometry in more detail and see that is not the case. 

In Figure \ref{Charged_BH_perturbed} (b), as we increase $t$, both the entanglement region (orange region) and Alice's entanglement wedge (red region) grow. Following \cite{Zhao:2017isy} we look at the time dependence of the spacetime volume of these regions. To simply calculations, we work in the following metric in Eddington-Finkelstein coordinates.
\begin{align*}
	ds^2 = -f(r)(du^*)^2+2du^*dr
\end{align*}
where $f(r) = \frac{(r-r_+)(r-r_-)}{l^2}$. The perturbation induces a shift in Kruskal coordinates by $\alpha = \frac{\Delta S}{S-S_0}e^{-\frac{2\pi}{\beta}t_w}$ where $\Delta S$ is the entropy brought in by the perturbation  \cite{Shenker:2013pqa}. 
%\begin{align*}
%	&\frac{1}{S}e^{\frac{2\pi}{\beta}(u^*-t_w)} = \frac{\tilde r_+-r}{r-r_-}\\
%	&1+\frac{1}{S}e^{\frac{2\pi}{\beta}(u^*-t_w)} = \frac{\tilde r_+-r_-}{r-r_-}\\
%	&\frac{\tilde r_+-r}{\tilde r_+-r_-} = \frac{\frac{1}{S}e^{\frac{2\pi}{\beta}(u^*-t_w)} }{1+}
%\end{align*}

%The spacetime volume element is given by $\frac{d\text{Vol}}{du^*} =  dr r^2 \text{Vol}(S^2)$.
We have
\begin{align}
\label{ratio}
	\frac{\frac{d\text{Vol}_{\text{red}}(t)}{dt}}{\frac{d\text{Vol}_{\text{orange}}(t)}{dt}+\frac{d\text{Vol}_{\text{red}}(t)}{dt}} =  \frac{\frac{\Delta S}{S-S_0}e^{\frac{2\pi}{\beta}(t-t_w)}}{1+\frac{\Delta S}{S-S_0}e^{\frac{2\pi}{\beta}(t-t_w)}}=\frac{e^{\frac{2\pi}{\beta}(t-t_w-t_*)}}{1+e^{\frac{2\pi}{\beta}(t-t_w-t_*)}}
\end{align}

With $\frac{S-S_0}{\Delta S}$ large, \eqref{ratio} is almost a step function. We see that after $t_w+t_*$ almost all of the increase of spacetime volume is in the red region. Despite the look of the Penrose diagram, the total spacetime volume of the orange region with Eddington-Finkelstein time larger than $t_w$ is finite. The orange region stops growing, which is consistent with the fact Alice's computations can no longer be undone by Bob after scrambling. The red region belonging to Alice grows linearly in time.

\bibliographystyle{JHEP}

\bibliography{main}

\end{document}